\documentclass[12pt]{article}
\usepackage{amssymb}
\usepackage{graphicx}
\graphicspath{ {./} }
\usepackage{slashed}
\usepackage[utf8]{inputenc}
\usepackage{amsmath}
\usepackage{bbold}
\usepackage{multirow}
\usepackage{tikz-feynman}

\newcommand{\rtwo}[1]{{\widetilde R}_{2,#1}}
\newcommand{\rtwobar}{{\widetilde R}_2}

\newcommand{\rtwohat}{{\widehat {\widetilde R}}_2}

\newcommand{\higgs}[1]{{\bf #1}_H}
\newcommand{\scalar}[1]{{\bf #1}_{S}}
\newcommand{\alphagut}{\alpha_{\mathrm{GUT}}}
\newcommand{\mgut}{M_{\mathrm{GUT}}}

\newcommand{\charge}{Q_{\mathrm{em}}}

\title{\textbf{Non-renormalizable grand unification utilizing the leptoquark mechanism of neutrino mass}}
\date{}
\author{Çağlar DOĞAN\footnote{e-mail:caglardogan80@gmail.com}$$}

\begin{document}
\maketitle

\begin{abstract}
We analyze a non-supersymmetric, non-renormalizable grand unified theory whose particle content is that of the Georgi-Glashow model augmented only by scalars from the \textbf{10} and \textbf{35} representations. A prediction of our model is a color sextet, weak isodoublet whose mass lies at 1 TeV (10 TeV) and that does not couple to Standard Model fermions at tree level. The leptoquark mechanism through which Majorana neutrinos radiatively acquire their masses relates the neutrino mass matrix to that of the down-type quarks. A consequence of this relation and perturbativity of coupling constants is the upper bound of $2.5 \times 10^{15}$ GeV on the masses of the scalar leptoquarks $S_1^*$ and ${\widetilde R}_{2}$. Electroweak mixing of these leptoquarks induces a B-L violating decay of the proton which indirectly contrains the mass of ${\widetilde R}_{2}$ to be greater than $2.1 \times 10^9$ GeV. We find the grand unification scale to exceed $1.4 \times 10^{16}$ GeV in all scenarios considered.
\end{abstract}

\section{Introduction}\label{intro}
Grand unified theories provide a rigorous framework with which to explore topics related to particle physics that the Standard Model (SM) fails to address. In its simplest form grand unification is embodied in the Georgi-Glashow (GG) model \cite{Georgi:1974sy}. This model doesn't just expose the anomaly free structure of the Standard Model (SM). It explains the quantization of charge, and correctly predicts the sine squared of the Weinberg angle and the ratio of the mass of the bottom quark to that of the tau lepton. Lastly, it predicts the -so far undetected- decay of the proton. \par
GG model, albeit elegant, has three major shortcomings. Firstly, its proposed unification of coupling constants is at odds with the observed values of physical parameters at the electroweak (EW) scale. Moreover, it fails to generate mass for neutrinos and hence it doesn't predict any mixing. Finally, there's disparity, at least for the lighter two generations, between its prediction for the ratio of the masses of charged leptons to down-type quarks and the measured values. \par
The solution to the first two problems requires addition of SU(5) multiplets that are not present in the original model. The third issue can likewise be tackled by the addition of a 45-plet to the GG model \`a la Georgi and Jarlskog \cite{Georgi:1979df}. In fact, this solves the first of the problems as well \cite{Dorsner:2006dj}. However, there's an alternative \cite{Ellis:1979fg} utilizing dimension five operators if one is to make a concession regarding the renormalizability of the theory. This latter route is the one we will take in this article. \par
In regard to generating small neutrino masses, the tree-level mechanisms fall under the three categories of Type-I \cite{Mohapatra:1979ia}, -II \cite{Minkowski:1977sc}, and -III \cite{Foot:1988aq} seesaw. The first and the third options involve fermions that are not in the SM, whereas in the second one scalar content of the SM is extended. These additional multiplets might as well remove the lack of unification -in the aforementioned sense- of the GG model. There are also recent nonrenormalizable \cite{Bajc:2006ia} and renormalizable \cite{FileviezPerez:2007bcw} hybrid models which utilize both type-I and -III seesaw. \par
As an alternative, loop models which naturally explain the smallness of neutrino mass due to loop effects were put forth. Early work had focused on radiative neutrino masses wherein the particles running in the loop were colorless. This is the case in the famous one-loop Zee model \cite{Zee:1980ai} and its construction within a grand unified framework in \cite{FileviezPerez:2016sal}. There are models that give mass to neutrinos at the two-loop level as well such as \cite{Babu:1988ki} and more recently \cite{Babu:2002uu,Babu:2010vp,Saad:2019vjo}. \par
There are phenomenological attempts \cite{FileviezPerez:2009ud,Law:2013dya,Klein:2019iws} at classification of all possible types of models generating neutrino masses at the loop level. \cite{Klein:2019jgb} tries to extend this to SU(5) GUTs. These investigations have revealed the possibility of particles running inside the loop having a nontrivial color quantum number. Perhaps more interestingly, they might even be leptoquarks. Leptoquarks carry both baryon and lepton number, and they might be scalar or vector particles. For an early work that employed leptoquarks check \cite{Murayama:1991ah}. For reviews of leptoquarks see \cite{Buchmuller:1986zs,Dorsner:2016wpm,Crivellin:2020mjs}, for constraints on their interactions see \cite{Buchmuller:1986iq,Davidson:1993qk}. An interesting mechanism dubbed the leptoquark mechanism originates precisely from this idea \cite{Nieves:1981tv,Mahanta:1999xd,AristizabalSierra:2007nf,Dorsner:2017wwn}. \par
As a matter of fact, leptoquark mechanism necessitates mixing between different types of leptoquarks. There's already one leptoquark, namely $S_1^*$ in the fundamental representation of $SU(5)$ in the GG model. A scalar transforming as the \textbf{10} of $SU(5)$, i.e., $\scalar{10}$ already contains the other desired leptoquark, $\rtwobar$. However, adding a $\scalar{10}$ to the particle content of the GG model by itself doesn't lead to a model that is compatible with experimental results regarding proton stability. As will be explained in greater detail in Section~\ref{unification}, a quick estimate at the one-loop level using Eq.(\ref{rationum2}) indicates that it actually takes at least 35 more scalars to achieve the appropriate grand unification (GUT) scale. \par
We have chosen to increase the GUT scale by introducing an additional scalar multiplet. Our criterion in doing so is such that there won't be any fermions in our model that are not already part of the SM. In addition to the $\scalar{10}$, we have opted for the smallest representation $\scalar{35}$\footnote{An earlier version of the manuscript used the $\scalar{40}$ representation, however, it entails additional complexity due to GUT scale mixing and the associated appearance of a GUT scale mixing angle in all results.} which will accomplish this objective. The leptoquark mechanism has already been discussed and outlined in reference \cite{Dorsner:2017wwn}. Since the authors of this reference are interested in building a renormalizable model, they discuss the $\scalar{45}$. Apart from this difference, to the best of the author's knowledge, this is the first time a GUT model that incorporates the leptoquark mechanism is constructed. This construction also allows us to refine the estimates made in the aforementioned article. \par
Regarding the outline of the article, after introducing the proposed model in Section~\ref{neutrino} we discuss the implementation of the leptoquark mechanism and obtain the equation relating the Majorana neutrino mass matrix to that of the down-type quarks. This is the major result of our article.  Section~\ref{unification} is devoted to constraints resulting from unification of coupling constants and the achievement of this unification at a high enough scale to avoid rapid proton decay. Mass matrices of charged fermions are given in Section~\ref{chargedfermion}. We conclude the article in Section~\ref{conclusion} by drawing conclusions and stating directions of future research based on what we have found.

\section{Neutrino Majorana mass matrix}\label{neutrino}
In the GG model \cite{Georgi:1974sy} SM fermions are neatly placed in the smallest nontrivial representations of $SU(5)$ as follows: $\overline{\bf 5}_{F,a} \equiv (d^C,L^T)_a=(\bar{3},1,1/3)\oplus (1,2,-1/2)$ and ${\bf 10}_{F,a} \equiv (u^C,Q,e^C)_a=(\bar{3},1,-2/3)\oplus (3,2,1/6)\oplus(1,1,1)$ where $a=1,2,3$ stands for the generation these fermions belong to. These are the left-handed fermions, the (right-handed) anti-particles of these are in the conjugate representations, whereas the SM Higgs field resides in the fundamental representation, i.e., ${\bf 5}_H \equiv (S_1^*,H)=(3,1,-1/3) \oplus (1,2,1/2)$. In addition, there is a (real) adjoint Higgs field ${\bf 24}_H \equiv (\Phi_8,\Phi_3,\Phi_{(3,2)},\Phi_{(\bar{3},2)},\Phi_{24})=(8,1,0)\oplus (1,3,0)\oplus(3,2,-5/6)\oplus (\bar{3},2,5/6)\oplus (1,1,0)$ which acquires a vacuum expectation value (VEV) in the SM singlet direction, that is $<{\bf 24}_H>= \text{diag}(2,2,2,-3,-3)v_{24}/\sqrt{30}$ after spontaneous symmetry breaking (SSB) at the GUT scale. As a result, gauge symmetry of the original theory is broken down to the SM subgroup, viz., $SU(5) \rightarrow SU(3)_C \times SU(2)_L \times U(1)_Y$. Charged fermions of the SM only become massive after the subsequent SSB that occurs at the electroweak (EW) scale due to the acquisition of a VEV by the neutral component of the SM Higgs field which breaks the symmetry further down to $SU(3)_C \times SU(2)_L \times U(1)_Y \rightarrow SU(3)_C \times U(1)_{\mathrm{em}}$. This VEV is taken to be $<{\bf 5}_H>^i=\delta^{i5}v_5/\sqrt{2}$ where $v_5=246$ GeV. For later use, we define the electromagnetic charge operator by $Q_{\mathrm{em}} \equiv T_3 + Y$ where $Y$ and $T_3$ are the hypercharge operator and third component of weak isospin operator, respectively. \par
In order to generate neutrino masses by the leptoquark mechanism in the most economical way in terms of the number of particles, we added to the GG model a scalar in the ${\bf 10}$ representation of SU(5), i.e., ${\bf 10}_S \equiv (\phi_1,\rtwobar,\phi_3)=(1,1,1) \oplus (3,2,1/6) \oplus ({\overline 3},1,-2/3)$. As was mentioned in the Introduction, this is not enough for the proton to have a long enough lifetime though. Therefore, we chose the other multiplet that we needed for this aim to be the scalar in the ${\bf 35}$ representation of SU(5). The decomposition of this multiplet under the SM gauge group is as follows: ${\bf 35}_S \equiv (\eta_1,\eta_2, \eta_3, \eta_4)=(1,4,3/2) \oplus (3,3,2/3) \oplus (6,2,-1/6) \oplus (10,1,-1)$. The above list demonstrates the presence of two leptoquarks in this model: $S_1^*$ and $\rtwobar$. Neutrinos will receive their masses through the mixing of leptoquarks $S_1^*$ and $\rtwobar$ after the electroweak SSB as we will see in this section. \par
In passing, we should state the well-known properties of the above SU(5) tensors. ${\bf 10}$ representation is anti-symmetric, i.e., ${\bf 10}^{ij}=-{\bf 10}^{ji}$ while the ${\bf 35}$ representation is completely symmetric in its indices, that is, ${\bf 35}^{ijk}={\bf 35}^{jik}={\bf 35}^{jki}$.
\subsection{EW mixing} \label{ewmixing}
\begin{figure}[h]
\centering
\includegraphics{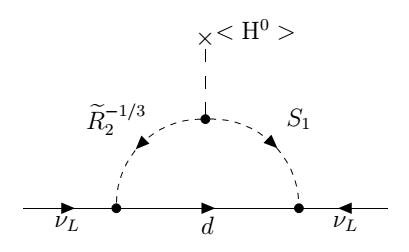}
\caption{Feynman diagram for Majorana neutrino mass}
\label{fig:neutrinomass}
\end{figure}

In this model, neutrinos radiatively receive their masses via the leptoquark mechanism in which the leptoquarks $\rtwobar$ and $S_1^*$ belonging to $\scalar{10}$ and $\higgs{5}$, respectively, run inside the loop shown in Figure \ref{fig:neutrinomass}. The Yukawa terms of SM matter fields,  ${\mathcal L}_{\text Y}$, is given in Eq.(\ref{yukawa}).
\begin{eqnarray} \label{yukawa}
\nonumber -{\mathcal L}_{\text Y} &=& Y^\nu_{ab} {\bf 10}_S^{ij} {\overline {\bf 5}}_{Fai} {\bf 5}^C_{Fbj} + Y^D_{ab} {\bf 10}^{ij}_{Fa} {\overline {\bf 5}}_{Fbi} {\bf 5}^*_{Hj} + \epsilon_{ijklm} {\bf 10}_{Fa}^{ij} Y^U_{ab} {\bf 10}_{Fb}^{kl} {\bf 5}_H^m \\
&+& \frac{1}{\Lambda} \left({\widetilde Y}^D_{ab} {\bf 10}^{ik}_{Fa} {\overline {\bf 5}}_{Fbi} {\bf 5}^*_{Hj} {\bf 24}_{Hk}^j +{\overline Y}^D_{ab} {\bf 10}^{kj}_{Fa} {\overline {\bf 5}}_{Fbi} {\bf 5}^*_{Hj} {\bf 24}_{Hk}^i + {\widetilde Y}^\nu_{ab} {\bf 10}_S^{ij} {\overline {\bf 5}}_{Fai} {\bf 5}^C_{Fbk} {\bf 24}_{Hj}^k \right) + H.c.
\end{eqnarray}
where $Y^\nu_{ab}=-Y^\nu_{ba}, Y^U_{ab}=Y^U_{ba}$ and subscripts $a,b=1,2,3$ of Yukawa coupling matrices denote generation indices. $Y^\nu$ can be taken to be an anti-symmetric complex matrix and $Y^U$ can be taken to be a symmetric complex matrix without loss of generality. The superscript $C$ in the terms involving $Y^\nu$ and ${\widetilde Y}^\nu$ denotes the charge conjugation operation. Although the third term in Eq.(\ref{yukawa}) which involves the u-type quarks is irrelevant for the leptoquark mechanism, we present it here for completeness. The Yukawa interactions of these leptoquarks and the mass term of the down-type quarks, which descend from the above Yukawa Lagrangian, that are both necessary to generate neutrino masses are given in Eq.(\ref{lqyukawa}).
\begin{eqnarray}
\label{lqyukawa}
\nonumber -{\mathcal L}_Y & \supset & \frac{1}{\sqrt{2}} \left(2Y^\nu - \frac{5v_{24}}{\Lambda\sqrt{30}} {\widetilde Y}^\nu\right) {\overline {d_R}}  L_L^\alpha {\widetilde R}_{2\alpha} - \frac{1}{\sqrt{2}} \left( Y^D + \frac{2v_{24}}{\Lambda\sqrt{30}} {\widetilde Y}^D -  \frac{3v_{24}}{\Lambda\sqrt{30}} {\overline Y}^D \right) {\overline {Q_L^c}}^\alpha \epsilon^{\alpha \beta}  L_L^\beta S_1 \\
&+& M^D ~ {\overline {d_R}} d_{L} + H.c.
\end{eqnarray}
Generation indices of the coupling constant matrices $Y^\nu, {\widetilde Y}^\nu,  Y^D, {\widetilde Y}^D$, and ${\overline Y}^D$ and the mass matrix $M^D$ of d-type quarks are suppressed. The  effective low-energy Lagrangian describing the interactions between scalar leptoquarks, quarks and leptons was written down by \cite{Buchmuller:1986zs} and contains the most general renormalizable Yukawa interactions that is consistent with the gauge symmetries of the SM. We give below only the relevant two terms of this effective Lagrangian:
\begin{eqnarray} \label{lqlq}
{\mathcal L}_{Y,LQ} \supset \lambda^{(R)}_{{\widetilde R}_{2}} {\overline d} P_L l {\widetilde R}_{2} + \lambda^{(L)}_{S_1} {\overline {q^c}} P_L i\tau_2 l S_1^L + H.c.
\end{eqnarray}
Regarding the symbols for the leptoquark fields we follow the notation of \cite{Dorsner:2016wpm}. $S_1^L~{\text{and}}~{\widetilde R}_2$ are the only two leptoquarks present in our model out of the seven appearing in the effective Lagrangian mentioned in the paragraph above. A comparison with Eq.(\ref{lqyukawa}) allows us to identify the coupling constants given in Eq.(\ref{lqlq}) with the ones we have introduced as follows: $\lambda^{(R)}_{{\widetilde R}_2}=-\frac{1}{\sqrt{2}} \left(2Y^\nu - \frac{5v_{24}}{\Lambda\sqrt{30}} {\widetilde Y}^\nu\right)$ and $\lambda^{(L)}_{S_1}= \frac{1}{\sqrt{2}} \left( Y^D + \frac{2v_{24}}{\Lambda\sqrt{30}} {\widetilde Y}^D -  \frac{3v_{24}}{\Lambda\sqrt{30}} {\overline Y}^D \right)$. \par
 The two-loop unification analysis that accounts for threshold corrections allows the identification of the mass, $M_V$, of the gauge bosons corresponding to broken symmetry generators after the GUT scale SSB given in Eq.(\ref{heavygaugeboson}) with the GUT scale, $\mgut$.
\begin{equation} \label{heavygaugeboson}
M_V = v_{24} \sqrt{\frac{5\pi \alphagut}{3}}~.
\end{equation}
The terms given hitherto aren't enough to give mass to neutrinos though, the crucial missing ingredient is the mixing between different leptoquark species. This mixing had been anticipated almost two decades ago by the authors of \cite{Hirsch:1996qy} due to the SSB that occurs at the EW scale and the most general renormalizable Lagrangian involving interactions of leptoquarks with the Higgs field had been written down. Here we give only the following SU(5) invariant interaction
\begin{equation} \label{nmscalar}
 \mathcal{L}_{\mathrm{mixed}} \supset \lambda_{5-10} {\bf 10}_S^{ij} {\bf 5}^*_{Hi} {\bf 5}^*_{Hk} {\bf 24}^k_{Hj} + H.c.
\end{equation}
from which the term given in Eq.(\ref{lqscalar}) coupling the component of $\rtwobar^*$ with $\charge=+1/3$ and $S_1^*$ is derived.
\begin{eqnarray} \label{lqscalar}
{\mathcal L}_{\mathrm{mixed}} & \supset & -h_{S_1}^{(L)} H i\tau_2 \rtwobar^*  S_1^* + H.c.
\end{eqnarray}
where $h_{S_1}^{(L)} =-v_{24} \lambda_{5-10} \sqrt{5/12}$. This is not the only term this low energy SSB leads to, more generally there's mixing between particles with electric charge $\charge=-1/3$ that are in the fundamental representation of the color gauge group, that is, particles whose quantum numbers are $(3,-1/3)$ after the EW SSB. These particles comprise $S_1^*, \rtwobar^{-1/3}$, and $\eta_2^{-1/3}$, where the superscript indicates the electric charge of the particle, from the ${\bf 5}_H, {\bf 10}_S,$ and ${\bf 35}_S$ representations, respectively. The Lagrangian, ${\mathcal L}_{\text{mixed}}$, which comprises the mixed terms among these scalar representations is given in Eq.(\ref{lmixed}). The first two terms in this Lagrangian lead to mixing between particles of electric charge $\charge=-1/3$. The absence of a term involving only the representations ${\bf 5}_H$ and ${\bf 35}_S$ in the following Lagrangian implies that $S_1^*$ and $\eta_2^{-1/3}$ don't mix at the same order as $S_1^*-\rtwobar^{-1/3}$, and $\rtwobar^{-1/3}-\eta_2^{-1/3}$ do. This is because a term of the form $\epsilon_{ijklm} {\bf 35}_S^{ijk} {\bf 5}_H^{l} {\bf 5}_H^{m}=0$ vanishes due to the complete symmetry of the ${\bf 35}_S$ representation! The first nonzero term ${\bf 35}_S^{ijk} {\bf 5}_{Hi}^* {\bf 5}_{Hj}^* {\bf 5}_{Hk}^*$ generates a mixing term of $\mathcal{O}(v_5^2)$, but this is much smaller than the mixing between the other two pairs of particles since those are of order $\mathcal{O}(v_5 v_{24})$. Therefore, for our purposes, mixing between $S_1^*$ and $\eta_2^{-1/3}$ can be ignored. We emphasize that this has no effect on our results, because its the mixing of $S_1^*$ and $\rtwobar^{-1/3}$ which generates the neutrino mass.
\begin{eqnarray} \label{lmixed}
\nonumber \mathcal{L}_{\text{mixed}} &=& \lambda_{5-10} {\bf 10}_S^{ij} {\bf 5}^*_{Hi} {\bf 5}^*_{Hk} {\bf 24}^k_{Hj} + \lambda_{10-35} {\bf 35}_S^{ijk} {\bf 10}^*_{Sil} {\bf 5}^*_{Hk} {\bf 24}_{Hj}^{l} + \epsilon_{ijklm} \left( g_{10-10} {\bf 10}_S^{ij} {\bf 10}_S^{kl} {\bf 5}_H^{m} \right. \\
&+& \left. \lambda_{10-10} {\bf 10}_S^{ij} {\bf 10}_S^{kl} {\bf 5}_H^{n} {\bf 24}_{Hn}^{m} + {\widetilde \lambda}_{10-10} {\bf 10}_S^{in} {\bf 10}_S^{kl} {\bf 5}_H^{m} {\bf 24}_{Hn}^{j} \right) + H.c.
\end{eqnarray}
Even though it doesn't affect the neutrino mass matrix, the above Lagrangian, nonetheless, is responsible for mixing between $\phi_3^*, \rtwobar^{+2/3}$, and $\eta_2^{+2/3}$ after the EW SSB, too. The details of the mixing between particles with electric charge $\charge=2/3$ are relegated to Appendix \ref{appendix:leptoquark}. \par
The mass eigenstates after EW SSB, namely ${\widehat S}_1^*, \rtwohat^{-1/3},$ and ${\widehat \eta_2^{-1/3}}$, are related through the rotation matrix, $R^{-1/3}$, to the interaction eigenstates which are $S_1^*, \rtwobar^{-1/3},$ and $\eta_2^{-1/3}$.
\begin{eqnarray} \label{rotation}
\begin{pmatrix} |{\widehat S}_1^*> \\ |\rtwohat^{-1/3}> \\ |{\widehat \eta_2^{-1/3}}> \end{pmatrix} & \equiv & R^{-1/3} \begin{pmatrix} |S_1^*> \\ |\rtwobar^{-1/3}> \\ |\eta_2^{-1/3}> \end{pmatrix}
\end{eqnarray}
The so-called squared mass matrix is diagonalized in the usual manner by the similarity transformation carried out using the matrix $R^{-1/3}$, that is, $\left({\mathcal M}^2_{-1/3}\right)_{\text{diag}}=R^{-1/3}~{\mathcal M}^2_{-1/3}~\left(R^{-1/3}\right)^T$. This transformation matrix is orthogonal since we take the couplings in the squared mass matrix, i.e., $\lambda_{5-10},~\text{and}~\lambda_{10-35}$, real without loss of generality. We write down only the elements on and above the diagonal because this is a symmetric matrix.
\begin{eqnarray} \label{massmatrix}
{\mathcal M}^2_{-1/3} &=& \begin{pmatrix}
m_{S_1^*}^2 & {\overline M}_{12}^2 & 0 \\ \cdot & m_{\rtwobar}^2 & {\overline M}_{23}^2 \\ \cdot & \cdot & m_{\eta_2}^2
\end{pmatrix}~.
\end{eqnarray}
Elements of the squared mass matrix given in Eq.(\ref{massmatrix}) are listed in terms of the quartic coupling constants $\lambda_{5-10}$, and $\lambda_{10-35}$ in Eqs.(\ref{massmatrixel}). There's a hierarchy between the diagonal elements and the off-diagonal ones of the squared mass matrix. The off-diagonal elements, assuming perturbative couplings, are much smaller than the differences of the squares of the masses appearing on the diagonal, i.e., ${\overline M}_{ij}^2 \sim v_5 v_{24} \ll \left|m_i^2-m_j^2\right|$ where  $i,j=1,2,3,4$ and $i \neq j$, provided that the masses aren't too close to each other. This hiearchy holds over the parameter space of our model that we explore in this article. 
\begin{eqnarray} \label{massmatrixel}
\nonumber {\overline M}_{12}^2 &=& -\frac{5}{2\sqrt{30}} v_5 v_{24} \lambda_{5-10}~, \\
{\overline M}_{23}^2 &=& -\frac{5}{6\sqrt{10}} v_5 v_{24} \lambda_{10-35}~.
\end{eqnarray}
We employ the hierarchy mentioned above to find the rotation matrix, $R^{-1/3}$, analytically up to corrections of ${\mathcal O}\left(\left({\overline M}_{ij}^2/\left(m_i^2-m_j^2\right)\right)^2\right)$ ($i \neq j$) as given below:
\begin{eqnarray} \label{rotmatrix}
R^{-1/3} = \begin{pmatrix} 1 & \frac{{\overline M}_{12}^2}{\left(m_{S_1^*}^2-m_{\rtwobar}^2\right)} & 0 \\ -\frac{{\overline M}_{12}^2}{\left(m_{S_1^*}^2-m_{\rtwobar}^2\right)} & 1 & \frac{{\overline M}_{23}^2}{\left(m_{\rtwobar}^2-m^2_{\eta_2}\right)} \\ 0 & -\frac{{\overline M}_{23}^2}{\left(m_{\rtwobar}^2-m^2_{\eta_2}\right)} & 1 \end{pmatrix}~.
\end{eqnarray}
We have checked numerically that the above approximation works at least to one part in a thousand accuracy in every scenario we consider in this article. The nonzero mixing angle in Eq.(\ref{rotmatrix}) proportional to ${\overline M}_{12}^2$ signal the generation of a Majorana mass term for neutrinos. The mixing angle in the rotation matrix, $R^{-1/3}$, proportional to ${\overline M}_{23}^2$ and the zero mixing angle are irrelevant for our discussion as the neutrino mass arises only from the mixing between $S_1^*$ and $\rtwobar^{-1/3}$. \par
Adapting the formula given in \cite{AristizabalSierra:2007nf} to our model, we obtain the equation for the neutrino Majorana mass matrix, ${\mathcal M}^\nu$, given below: 
\begin{eqnarray} \label{majoranamass1}
\nonumber \left({\mathcal M}^\nu\right)_{i i^\prime} &=& -\frac{3}{32\pi^2} \sum_{\begin{array}{c} j=1,2,3 \\ k,l=d,s,b \end{array}} \left(M^D\right)_{kl} \left[ \frac{m_k^2 \text{log}\left(m_k^2\right)-m_j^2 \text{log}\left(m_j^2\right)}{\left( m_k^2-m^2_j\right)} \right] R^{-1/3}_{j1} R_{j2}^{-1/3} \\
\nonumber && \times \left[ \left( 2Y^\nu - \frac{5v_{24}}{\Lambda\sqrt{30}} {\widetilde Y}^\nu\right)_{ik} \left( Y^D + \frac{2v_{24}}{\Lambda\sqrt{30}} {\widetilde Y}^D -  \frac{3v_{24}}{\Lambda\sqrt{30}} {\overline Y}^D \right)_{i^\prime l} \right. \\
&& + \left. \left( 2Y^\nu - \frac{5v_{24}}{\Lambda\sqrt{30}} {\widetilde Y}^\nu\right)_{i^\prime l} \left( Y^D + \frac{2v_{24}}{\Lambda\sqrt{30}} {\widetilde Y}^D -  \frac{3v_{24}}{\Lambda\sqrt{30}} {\overline Y}^D \right)_{ik} \right]~. 
\end{eqnarray}
The formula in the above reference was given in the mass eigenstate basis of d-type quarks wherein $M^D_{kl}=m_k \delta_{kl}, k=d,s,b$ , so that we recover their formula. However, we will work in the Pontecorvo-Maki-Nakagawa-Sakata (PMNS) basis in which the mass matrix of charged leptons is diagonal, i.e., $M^E_{kl}=m_k \delta_{kl}, k=e,\mu,\tau$, whereas that of d-type quarks is not and is equal to $M^D=M^D_{\mathrm{diag}} V^\dagger_{\mathrm{CKM}}$. To simplify Eq.(\ref{majoranamass1}) we use the orthogonality relation $\left(R^{-1/3}\right)_{ij} \left(R^{-1/3}\right)_{kj}=\delta_{ik}$. This step would still have been valid, if the coupling constants had been complex.
\begin{eqnarray} \label{majoranamass2}
\nonumber \left({\mathcal M}^\nu\right)_{i i^\prime} &=& -\frac{3}{32\pi^2} \sum_{k=d,s,b} \left(M^D\right)_{kl} \left[ \left( 2Y^\nu - \frac{5v_{24}}{\Lambda\sqrt{30}} {\widetilde Y}^\nu\right)_{ik} \left( Y^D + \frac{2v_{24}}{\Lambda\sqrt{30}} {\widetilde Y}^D -  \frac{3v_{24}}{\Lambda\sqrt{30}} {\overline Y}^D \right)_{i^\prime l} \right. \\
\nonumber && + \left. \left( 2Y^\nu - \frac{5v_{24}}{\Lambda\sqrt{30}} {\widetilde Y}^\nu\right)_{i^\prime l} \left( Y^D + \frac{2v_{24}}{\Lambda\sqrt{30}} {\widetilde Y}^D -  \frac{3v_{24}}{\Lambda\sqrt{30}} {\overline Y}^D \right)_{ik} \right] \\
& \times & R_{12}^{-1/3} \left[ \frac{m_{S_1^*}^2 \text{log}\left(m_k^2/m_{S_1^*}^2\right)}{\left( m_k^2-m_{S_1^*}^2\right)} - \frac{m_{\rtwobar}^2 \text{log}\left(m_k^2/m_{\rtwobar}^2\right)}{\left( m_k^2-m_{\rtwobar}^2\right)} \right] ~.
\end{eqnarray}
At this point, we make the approximation of ignoring the masses of the down-type quarks in the denominators to obtain Eq.(\ref{majoranamass3}). This is well justified, because even for the lowest values of the masses of leptoquarks $S_1^*$ and $\rtwobar$ in this article, the relation $m_b = 4.18~\text{GeV}~\ll m_{S_1^*} = 2.8 \times 10^{11}~\text{GeV},~m_{\rtwobar} = 2.4 \times 10^{9}~\text{GeV}$ holds.
\begin{eqnarray} \label{majoranamass3}
\nonumber {\mathcal M}^\nu &\simeq& \frac{3}{32\pi^2} \left(\frac{1}{2} \sqrt{\frac{5}{6}} v_5 v_{24} \lambda_{5-10} \right) \left[ \left( 2Y^\nu - \frac{5v_{24}}{\Lambda\sqrt{30}} {\widetilde Y}^\nu\right) M^D \left( Y^D + \frac{2v_{24}}{\Lambda\sqrt{30}} {\widetilde Y}^D -  \frac{3v_{24}}{\Lambda\sqrt{30}} {\overline Y}^D \right)^T \right. \\
&& + \left. \left( Y^D + \frac{2v_{24}}{\Lambda\sqrt{30}} {\widetilde Y}^D -  \frac{3v_{24}}{\Lambda\sqrt{30}} {\overline Y}^D \right) M^D \left( 2Y^\nu - \frac{5v_{24}}{\Lambda\sqrt{30}} {\widetilde Y}^\nu\right)^T \right] \frac{\text{log}\left(m_{S_1^*}^2/m_{\rtwobar}^2\right)}{\left(m_{S_1^*}^2-m_{\rtwobar}^2\right)}
\end{eqnarray}
where the matrix $M^D$ has elements $M^D_{ij}=m_i V^*_{\mathrm{CKM},ji}$ for $i,j=d,s,b$ in Eq.(\ref{majoranamass3}). Therefore, we arrive at the major formula of the article.
\begin{eqnarray} \label{majorformula}
\nonumber {\mathcal M}^\nu &=&  \frac{3v_5 \lambda_{5-10}}{64\pi^2\sqrt{2\pi}} \frac{\mgut}{\sqrt{\alphagut}} \frac{\text{log}\left(m_{S_1^*}^2/m_{\rtwobar}^2\right)}{\left(m_{S_1^*}^2-m_{\rtwobar}^2\right)} \\
\nonumber && \times \left[ \left( 2Y^\nu - \frac{5v_{24}}{\Lambda\sqrt{30}} {\widetilde Y}^\nu\right) M^D \left( Y^D + \frac{2v_{24}}{\Lambda\sqrt{30}} {\widetilde Y}^D -  \frac{3v_{24}}{\Lambda\sqrt{30}} {\overline Y}^D \right)^T \right. \\
&& + \left. \left( Y^D + \frac{2v_{24}}{\Lambda\sqrt{30}} {\widetilde Y}^D -  \frac{3v_{24}}{\Lambda\sqrt{30}} {\overline Y}^D \right) M^D \left( 2Y^\nu - \frac{5v_{24}}{\Lambda\sqrt{30}} {\widetilde Y}^\nu\right)^T \right]
\end{eqnarray}
Now, we use Eq.(\ref{majorformula}) to obtain upper bounds on the masses of $S_1^*$ and $\rtwobar$. First, we observe that $M^D_{kl} \simeq m_b \left[ \delta_{k3} \delta_{kl} + \mathcal{O}(m_s V_{ts}^*/m_b V_{tb} \sim 10^{-3}, V_{cb}^*/V_{tb} \sim 4 \times 10^{-2} )\right]$. To this end, we assume that there's no hierarchy between elements of the Yukawa coupling matrices $Y^\nu$ and $Y^D$ belonging to different generations, that is, $Y^\nu_{i1} \sim Y^\nu_{i2} \sim Y^\nu_{i3}$ and $Y^D_{i1} \sim Y^D_{i2} \sim Y^D_{i3}$ so that the sum is dominated by the contribution of the bottom quark! We use the standard parametrization of the $U_{\mathrm{PMNS}}$ matrix which diagonalizes ${\mathcal M}^\nu$, that is, ${\mathcal M}^\nu=U_{\mathrm{PMNS}} {\mathcal M}^\nu_{\mathrm{diag}} U_{\mathrm{PMNS}}^T$.
\begin{eqnarray}
U_{\mathrm{PMNS}} = \begin{pmatrix} c_{12} c_{13} & s_{12}c_{13} & s_{13}e^{-i\delta} \\ -s_{12}c_{23}-c_{12}s_{23}s_{13}e^{i\delta} & c_{12}c_{23}-s_{12}s_{23}s_{13}e^{i\delta} & s_{23}c_{13} \\ s_{12}s_{23}-c_{12}c_{23}s_{13}e^{i\delta} & -c_{12}s_{23}-s_{12}c_{23}s_{13}e^{i\delta} & c_{13}c_{23} \end{pmatrix} \times \mathrm{diag} \left(1,e^{i\eta_2/2},e^{i\eta_3/2} \right)
\end{eqnarray}
where $c_{ij} \equiv \mathrm{cos}\theta_{ij}$ and $s_{ij} \equiv \mathrm{sin}\theta_{ij}$, and the (central values of the) experimentally measured values of neutrino mass squared differences $m_{\nu_1} = 0$, $m_{\nu_2} = \sqrt{\Delta m^2_{\mathrm{sol}}}$, $m_{\nu_3} = \sqrt{\Delta m^2_{\mathrm{atm}}+\Delta m^2_{\mathrm{sol}}}$ for normal ordering (NO); $m_{\nu_3} = 0$, $m_{\nu_2} = \sqrt{\Delta m^2_{\mathrm{atm}}}$, $m_{\nu_1} = \sqrt{\Delta m^2_{\mathrm{atm}}-\Delta m^2_{\mathrm{sol}}}$ for inverted ordering (IO) where $\Delta m^2_{\mathrm{sol}} \equiv \Delta m^2_{\mathrm{21}}=(7.53 \pm 0.18) \times 10^{-5}$ eV$^2$, $\Delta m^2_{\mathrm{atm}} \equiv \left| \Delta m^2_{\mathrm{32}}\right| = 2.437~(2.519) \pm 0.033 \times 10^{-3}$ eV$^2$, and the mixing angles sin$^2\theta_{12}=0.307 \pm 0.013$, sin$^2\theta_{23}=0.547^{+0.018}_{-0.024} ~(0.534^{+0.021}_{-0.024} )$, sin$^2\theta_{13}=(2.20 \pm 0.07)\times 10^{-2}$ and phase $\delta=(1.23 \pm 0.21) \pi$ [rad]. Taking the CP violating phase to be $\delta=\pi$ and the Majorana phases to be zero $\eta_2=\eta_3=0$ ($\eta_2=\eta_1=0$) in NO (IO) for simplicity, we find the smallest absolute value of entries to be 0.0024 (0.0075) eV for normal (inverted) hierarchy of neutrino masses and the largest entry to be 0.0298 (0.0447) eV for a normal (inverted) hiearchical neutrino spectrum. \par
As can be seen in Eq.(\ref{majorformula}), $\left({\mathcal M}^\nu\right)_{i i^\prime} \propto \left(Y^\nu_{i3} Y^D_{i^\prime 3}+Y^\nu_{i^\prime 3} Y^D_{i3}\right)/\left(m_{S_1^*}^2-m_{\rtwobar}^2\right)$, therefore, assuming perturbative coupling constants, one finds that the leptoquark masses are limited from above, i.e., $m_{S_1^*},~m_{\rtwobar} \leq 2.5 \times 10^{15}$ GeV. If the masses of the leptoquarks are comparable to the upper bound, it is dubbed the heavy leptoquark regime in reference \cite{Dorsner:2017wwn}. Our model can accomodate both types of hierarchical neutrino spectrum as we use the more stringent value of 0.0447 eV in obtaining the upper bound. Of course, this bound can be evaded if the leptoquark masses are sufficiently close to each other, however, we don't consider such isolated regions of parameter space. Although this upper bound on the masses of scalar leptoquarks was anticipated in reference \cite{Dorsner:2017wwn}, it's slightly larger than the value of $5 \times 10^{13}$ GeV estimated there. Our refinement of the estimate given in that reference can be attributed to using $m_b=4.18$ GeV, $v_5=246$ GeV, and multiplying by $\mgut=1.42 \times 10^{16}$ GeV and $\sqrt{\alphagut^{-1}}=\sqrt{35}$ in our calculation. \par
\begin{figure}[h]
\centering
\includegraphics{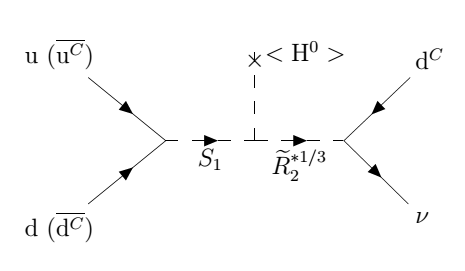}
\caption{B-L violating s channel contribution to proton decay induced by the EW mixing between leptoquarks $S_1^*$ and ${\widetilde R}_{2}^{-1/3}$. Spectator u quark is not shown in the figure.}
\label{fig:proton decay}
\end{figure}

In regard to the lower bounds on the masses of scalar leptoquarks, we will see in Section~\ref{unification} that the mass of the leptoquark $S_1^*$ is bounded from below due to proton decay contraints. On the other hand, $\rtwobar$ by itself doesn't lead to proton decay at tree level and thus can seemingly be as light as the Z boson, because it doesn't have diquark couplings. However, as a consequence of EW mixing between these two leptoquark species a new contribution to proton decay appears as illustrated in Figure~\ref{fig:proton decay}.   
This is a s-channel decay which violates the difference of baryon and lepton numbers, $B-L$, and it leads to a partial decay rate that is proportional to the squares of the propagators of the two leptoquarks which in the limit $s \ll m^2_{\rtwobar}, m^2_{S_1^*}$ implies a decay rate scaling as the fourth power of the inverse of the mass of each leptoquark. Under the assumption that this is the dominant contribution to proton decay, the lifetime of the proton which is inversely proportional to the decay rate, then, scales as the fourth power of the product of the masses of the two leptoquarks. This is clearly seen to be the case in Eq.(\ref{decayest}):
\begin{eqnarray} \label{decayest}
\nonumber && \left( 1.95 \times 10^{-64} ~\mathrm{GeV}^{-6} \right) \frac{\left( m_{S_1^*} m_{\rtwobar} \right)^4}{\left( {\overline M}_{12}^2/v_5 \right)^2} \\
&& > \left[ 0.19 \left| C\left(\nu_i,s,d^C \right) + C\left(\nu_i,s^C,d^C \right)\right|^2 + 2.49 \left| C\left(\nu_i,d,S^C \right) + C\left(\nu_i,d^C,s^C \right)\right|^2 \right]~.
\end{eqnarray}
which is taken from \cite{Dorsner:2005fq} and adapted to this model. The presence of $M_{12}^2$ in the denominator signifies that the diagram originates from mixing between the leptoquarks and would lead to an infinite proton lifetime unless the leptoquarks mix. According to the same reference \cite{Dorsner:2005fq}, the terms on the right side of the equation are of order $\mathcal{O}(10^{-4})$ which implies that the product of the masses of the two leptoquarks must exceed $2.0 \times 10^{23}$ GeV$^2$. For the scenario in which $m_{\rtwobar} < m_{S_1^*} = 9.2 \times 10^{13}$ GeV this sets a limit on the mass of $\rtwobar$ from below as $m_{\rtwobar} > 2.1 \times 10^9$ GeV.

\section{Unification in the proposed model}\label{unification}
The lynchpin of a grand unified theory is the unification of coupling constants at the GUT scale, $\mgut$. In order to analyze this unification we need to take into account the running of the fine structure constants, $\alpha_i(\mu)$, as the renormalization scale changes. To simplify the discussion we will consider only the one-loop beta functions, however, our results will, of course, be obtained by implementing the running of fine structure constants up to two-loop order. In Eq.(\ref{running}), we give the variation of the fine structure constants resulting from the one-loop beta functions.
\begin{equation}\label{running}
\alpha^{-1}_i(m_Z) = \alpha^{-1}_{\mathrm{GUT}} + \frac{1}{2\pi} b_i ~ \text{log}\left(\frac{\mgut}{m_Z}\right)
\end{equation}
for $i=1,2,3$ corresponding to the SM interactions. Here $\alphagut \equiv \alpha_i(\mgut)$ represents the common value of the fine structure constants at the GUT scale. The SM values of the one-loop beta function coefficients $b_i$ are given below:
\begin{equation}
b_1=\frac{41}{10}~,~~~~ b_2=-\frac{19}{6}~,~~~~ b_3=-7 ~.
\end{equation}
The above expressions for the $b_1,$ and $b_2$ coefficients include the contribution of the SM Higgs doublet. However, SM particles do not lead to unification of coupling constants. Therefore, new particles with masses $M_I$ above the weak scale $m_Z$ and below the GUT scale, and beta function coefficients $b_{iI}$ have to contribute to the running to ensure unification of coupling constants. The effect of these hypothetical particles is to replace  as shown below   
\begin{equation}
B_i = b_i + \sum_I b_{iI} r_I ~~ \text{where} ~~ r_I \equiv \frac{\text{log}\left(\mgut/M_I\right)}{\text{log}\left(\mgut/m_Z\right)}
\end{equation}
the one-loop beta function coefficients in Eq.(\ref{running}) with effective coefficients $B_i$. Moreover, the equations that need to be solved can be reduced to two by taking the differences of those appearing in Eq.(\ref{running}) and defining new variables $B_{ij} \equiv B_i-B_j$. It was realized decades ago by the authors of \cite{Giveon:1991zm} that regardless of the details of the GUT under consideration there are relations which must be satisfied relating the measured values of physical observables at the weak scale to the ratio of the effective coefficients and of the GUT scale to the weak scale as follows:
\begin{eqnarray}
\label{ratio1}
\frac{B_{23}}{B_{12}} = \frac{5}{8} \cdot \frac{\text{sin}^2\theta_W-\alpha_{\text{em}}/\alpha_s}{3/8-\text{sin}^2\theta_W} ~, \\
\label{ratio2}
\text{log}\left(\frac{\mgut}{m_Z}\right) = \frac{16\pi}{5\alpha_{\text{em}}} \cdot \frac{3/8-\text{sin}^2\theta_W}{B_{12}} ~. 
\end{eqnarray}
When the physical observables in Eqs.(\ref{ratio1}) and (\ref{ratio2}) are replaced by their current values in the $\overline{\text{MS}}$ scheme \cite{ParticleDataGroup:2022pth} sin$^2\theta_W(m_Z)=0.23120 \pm 0.00015,~\alpha_{\text{em}}^{-1}(m_Z)=127.906 \pm 0.019$ and $\alpha_s(m_Z)=0.1187 \pm 0.002$ one obtains the following numerical values of these ratios:
\begin{eqnarray}
\label{rationum1}
\frac{B_{23}}{B_{12}}=0.719 \pm 0.005 ~, \\
\label{rationum2}
\text{log}\left(\frac{\mgut}{m_Z}\right) = \frac{184.9 \pm 0.2}{B_{12}} ~.
\end{eqnarray}
The ratio of the two coefficients, which is $B_{23}/B_{12} \approx 0.53$ for the SM, is 25\% less than what it has to be for successful unification. In order to make a statement about the required value of $B_{12}$ one needs both to know the experimental lower bound and to have a theoretical prediction for the lifetime of the proton. \par 
Let us  begin by reviewing the current status of experimental results. The Super-Kamiokande experiment based on the absence of  a signal for proton decays to a neutral pion and a positron, and to a positively charged kaon and an anti-neutrino has so far set the lower limit of the proton lifetime to $\tau_p (p \rightarrow \pi^0 e^+) > 2.4 \times 10^{34}$ years \cite{Super-Kamiokande:2020wjk}, and $\tau_p (p \rightarrow K^+ {\overline \nu}) > 5.9 \times 10^{33}$ years \cite{Super-Kamiokande:2014otb}. The Hyper-Kamiokande collaboration is expected to raise this to $\tau_p (p \rightarrow \pi^0 e^+ ) > 6.3 \times 10^{34}$ years \cite{Hyper-Kamiokande:2018ofw}. On the theoretical side, there are upper bounds on partial decay rates \cite{Nath:2006ut} for a particular decay channel due to gauge boson exchange that can then be converted into a lower bound for the lifetime of the proton. Here we assume the gauge boson mediated decay processes to be dominant over those mediated by scalar leptoquarks. \par 
Using the upper bounds on the nucleon lifetime due to partial decays given in Eqs.(\ref{lifetime}) one can find a lower bound on the GUT scale. These equations are obtained from those in \cite{Dorsner:2005fq} by setting $k_2=0$.
\begin{eqnarray} \label{lifetime}
\nonumber \Gamma \left(p \rightarrow \pi^0 e_\beta^+\right) & \leq & \frac{5m_p}{16\pi f_\pi^2} A_L^2 \left| \alpha \right|^2 \left(1+D+F\right)^2 k_1^4, \\
\Gamma \left(p \rightarrow K^+ {\overline \nu} \right) &\leq& \frac{\left(m_p^2-m^2_{K}\right)^2}{8\pi f_\pi^2 m_p^3} A_L^2 \left| \alpha \right|^2 \left\{ \left(\frac{2m_pD}{3m_B}\right)^2+\left[1+\frac{m_p}{3m_B}\left(D+3F\right)\right]^2 \right\} k_1^4
\end{eqnarray}
where $m_p=938.3 ~\text{MeV}, m_{K^+}=493.7 ~\text{MeV}, D=0.81, F=0.44, m_B=1150 ~\text{MeV}, f_\pi=139 ~\text{MeV}, A_L=1.43$, and the most conservative value $\alpha=0.003 ~\text{GeV}^3$. The variable $k_1$ in Eq.(\ref{lifetime}) is defined as $k_1^2 \equiv 4\pi \alpha_{\text{GUT}}/M^2_{\text{GUT}}$. \par
This lower bound on the GUT scale is $M_{\text{GUT}} \geq  1.45 \times 10^{16} \sqrt{\alpha_{\text{GUT}}} \sqrt{\alpha/0.003 ~ {\text GeV}^3}$ in terms of $\alpha_{\text{GUT}}$ and the nucleon matrix element which we take to be $\alpha = 0.015 ~ {\text GeV}^3$ \cite{Yoo:2021gql}. In obtaining this expression we have identified the mass of the gauge bosons $(3,2,-5/6) \oplus ({\overline 3},2,5/6)$ that are responsible for proton decay with the GUT scale, i.e., $M_V = M_{\text{GUT}}$. In all scenarios we consider in this article the value of the coupling constant at the unification scale is approximately $\alpha_{\text{GUT}} \approx 1/35$. Therefore, in order for the model presented here to be compatible with proton decay experiments the GUT scale we find should exceed $M_{\text{GUT}} \geq 5.5 \times 10^{15}$ GeV. As can be seen in Tables \ref{table:masses1} and \ref{table:masses2}, this condition is met in all scenarios. \par
By looking at Eq.(\ref{rationum2}) we see that the aforementioned lower bound on the GUT scale implies an upper bound on the value of $B_{12} \leq 5.83$ which is substantially lower than $B_{12}^{\mathrm{SM}}=109/15 \approx 7.27$ of the SM. We proved in Section~\ref{neutrino} that $m_{\rtwobar} \geq 2.1 \times 10^9$ GeV, therefore the contribution of a single $\scalar{10}$ is $\Delta B_{12} \leq 0.22$ which means that the remaining scalar representations must supply at least $7.3-5.83-0.22=1.25$ to $B_{12}$. If one just uses $\scalar{5}$ it takes nineteen of them, with only $\scalar{24}$ it requires more than three of them. We could use only $\scalar{15}$ which contributes $\Delta B_{12} \leq 0.29$, hence at least five of them would be necessary, with only $\scalar{10}$ this figure would go up to six. Thus, adding a single $\scalar{35}$ representation to the already existing particles provides the minimal number of particles required for unification at a GUT scale that is compatible with experimental results on the proton lifetime. 
\par 
When a given SU(5) representation is degenerate, i.e., all the multiplets comprising such a representation have a common mass, it doesn't affect the value of $B_{12}$ and thus doesn't raise the unification scale above the lower bound given above. Therefore, the split multiplets of the representations we have added to the GG model must be non-degenerate. This splitting is accomplished through the mass terms generated by contractions of the representations ${\bf 10}_S$, ${\bf 35}_S$ and adjoint scalars in ${\bf 24}_H$ given in Eqs.(\ref{mass10}) and (\ref{mass35}). We state the mass terms of the $\scalar{10}$ representation in Eq.(\ref{mass10}),
\begin{eqnarray} \label{mass10}
\nonumber {\mathcal L}_{\text{mass}} & \supset & -m_{10}^2 {\bf 10}_S^{ij} {\bf {10}}^*_{Sij} - g_{10} {\bf {10}}^{ij}_S {\bf 24}_{Hj}^k {\bf 10}^*_{Ski} - \lambda_{10} {\bf 10}^{ij}_S {\bf 24}_{Hj}^k {\bf 24}_{Hi}^l {\bf 10}^*_{Slk} \\
&& -\lambda_{g,10} {\bf 10}^{ij}_S {\bf 24}_{Hj}^k {\bf 24}_{Hk}^l {\bf 10}^*_{Sli} - \lambda_{m^2,10} {\bf 10}^{ij}_S {\bf 10}_{Sij}^* {\bf 24}^k_{Hl} {\bf 24}^l_{Hk}~.
\end{eqnarray}
One can derive the masses of the split multiplets of the $\scalar{10}$ representation given below using the mass terms listed in Eq.(\ref{mass10}),
\begin{eqnarray} \label{appmass10}
\nonumber m_{\phi_1}^2 &=& M_{10}^2 - \frac{3}{\sqrt{30}} g_{10}v_{24} + \frac{9}{30} \lambda_{g,10}v_{24}^2 + \frac{9}{30} \lambda_{10} v_{24}^2~,\\
\nonumber m_{\rtwobar}^2 &=& M_{10}^2 - \frac{1}{2\sqrt{30}} g_{10}v_{24} + \frac{13}{60} \lambda_{g,10}v_{24}^2 - \frac{6}{30} \lambda_{10}v_{24}^2~,\\
m_{\phi_3}^2 &=& M_{10}^2 +\frac{2}{\sqrt{30}} g_{10}v_{24} + \frac{4}{30} \lambda_{g,10}v_{24}^2 + \frac{4}{30} \lambda_{10}v_{24}^2
\end{eqnarray}
where $M_{10}^2 \equiv m_{10}^2 + v_{24}^2 \lambda_{m^2,10}$. The independent coupling constants $M^2_{10},~g_{10},~\lambda_{g,10},$ and $\lambda_{10}$ in Eq.(\ref{appmass10}) provide enough parameters to arrange the masses of the split multiplets in $\scalar{10}$ as is necessary for unification.\par
The mass terms of the $\scalar{35}$ representation are given in Eq.(\ref{mass35}),
\begin{eqnarray} \label{mass35}
\nonumber {\mathcal L}_{\text{mass}} & \supset & -m_{35}^2 {\bf 35}^{ijk}_S {\bf 35}^*_{Sijk} - g_{35} {\bf 35}^{ijk}_S {\bf 24}^l_{Hk} {\bf 35}^*_{Sijl} -\lambda_{35} {\bf 35}^{ijk}_S {\bf 24}^l_{Hj} {\bf 24}^m_{Hk} {\bf 35}^*_{Silm} \\
&& - \lambda_{m^2,35} {\bf 35}^{ijk}_S {\bf 35}_{Sijk}^* {\bf 24}^l_{Hm} {\bf 24}^m_{Hl} - \lambda_{g,35} {\bf 35}^{ijk}_S {\bf 24}^l_{Hm} {\bf 24}^m_{Hk} {\bf 35}^*_{Sijl}~.
\end{eqnarray}
Similarly, the masses of the split multiplets comprising the $\scalar{35}$ representation given below follow from the mass terms stated in Eq.(\ref{mass35}):
\begin{eqnarray} \label{appmass35}
\nonumber m_{\eta_1}^2 &=& M_{35}^2 + \frac{3}{10}\lambda_{35}v_{24}^2 +\frac{3}{10} \lambda_{g,35}v_{24}^2 - \frac{\sqrt{3}}{\sqrt{10}} g_{35}v_{24}~,\\
\nonumber m_{\eta_2}^2 &=& M_{35}^2 - \frac{1}{30} \lambda_{35}v_{24}^2 +\frac{11}{45}\lambda_{g,35}v_{24}^2 - \frac{2\sqrt{2}}{3\sqrt{15}} g_{35}v_{24}~,\\
m_{\eta_3}^2 &=& M_{35}^2 - \frac{4}{45}\lambda_{35}v_{24}^2 +\frac{17}{90}\lambda_{g,35}v_{24}^2 + \frac{1}{3\sqrt{30}} g_{35}v_{24}
\end{eqnarray}
where $M_{35}^2 \equiv m_{35}^2 + v_{24}^2 \lambda_{m^2,35}$~. The independent coupling constants $M^2_{35},~\lambda_{35},~\lambda_{g,35}$ and $g_{35}$ in Eq.(\ref{appmass35}) allow the masses of only three of the four split multiplets to be varied freely. The mass of the remaining split multiplet, $m_{\eta_4}$, is determined in terms of the masses of the other split multiplets by the following equation \cite{Dorsner:2019vgf}:
\begin{eqnarray} \label{constraint35}
m_{\eta_4}^2 &=& m_{\eta_1}^2 - 3m_{\eta_2}^2 + 3m_{\eta_3}^2~.
\end{eqnarray}
Both $\eta_1$ and $\eta_4$ are decoupled from the theory below the GUT scale. Since the renormalizable terms provide enough parameters already in the case of $\scalar{10}$ and since the constraint between the masses of the split multiplets of $\scalar{35}$ in Eq.(\ref{constraint35}) is not very restrictive, we have not listed the dimension 5 operators giving mass to the $\scalar{10}$ and $\scalar{35}$ that must be included for consistency. Regarding the multiplets of the $\higgs{5}$, it's well-known that while the SM Higgs doublet $H$ is at the EW scale, the triplet, i.e., $S_1^*$ can be made as heavy as desired. This is the subject of the notorious doublet-triplet splitting problem. \par 
There are also terms related to the Higgs scalar potential \cite{Dorsner:2005fq}, ${\mathcal L}_{\text{Higgs}}$,
\begin{eqnarray} \label{higgspot}
\nonumber {\mathcal L}_{\text{Higgs}} &=& \frac{{\mu_{24}}^2}{2} {\bf 24}^i_{Hj} {\bf 24}^j_{Hi} - \frac{a_{24}}{4} \left({\bf 24}^i_{Hj} {\bf 24}^j_{Hi}\right)^2 - \frac{b_{24}}{2} {\bf 24}^i_{Hj} {\bf 24}^j_{Hk} {\bf 24}^k_{Hl} {\bf 24}^l_{Hi} - \frac{c_{24}}{3} {\bf 24}^i_{Hj} {\bf 24}^j_{Hk} {\bf 24}^k_{Hi} \\
\nonumber &+& \frac{\mu_5^2}{2} {\bf 5}_H^i {\bf 5}^*_{Hi} - \frac{a_5^2}{4} \left( {\bf 5}_H^i {\bf 5}^*_{Hi} \right)^2 - g_{5-24} {\bf 5}_H^i {\bf 5}^*_{Hj} {\bf 24}_{Hi}^j - \lambda_{\mu_5^2} {\bf 5}_H^i {\bf 5}^*_{Hi} {\bf 24}^j_{Hk} {\bf 24}^k_{Hj} \\
&-& \lambda_{g,5-24} {\bf 5}_H^i {\bf 5}^*_{Hk} {\bf 24}^j_{Hi} {\bf 24}^k_{Hj}~.
\end{eqnarray}
GG model has a $\higgs{24} \rightarrow -\higgs{24}$ discrete symmetry that is not respected by the operators of dimension-5 that appear among the terms of ${\mathcal L}_Y$ in Eq.(\ref{yukawa}). This justifies our inclusion of terms that involve a single power of $\higgs{24}$ among the mass terms, too (Some of the non-renormalizable terms not listed in ${\mathcal L}_{\text{mass}}$ violate this symmetry as well). Consistency, then, requires that a cubic term appear in the first line of ${\mathcal L}_{\text{Higgs}}$ as shown above in Eq.(\ref{higgspot}). This term invalidates the relation $m_{\Phi_3}^2=4m_{\Phi_8}^2$ \cite{Buras:1977yy} between the masses of isotriplet scalars and scalar gluons. Consequently, the masses of these split multiplets can also be adjusted as desired. We will not make use of this fact, though, as these split multiplets are neither essential for meeting unification constraints nor do they play any role in generation of masses for neutrinos via the leptoquark mixing mechanism. \par
\begin{table}
\centering
\caption{$b_{iI}-b_{jI}$ coefficients of the split multiplets in the proposed model}
\label{table:beta}
\begin{tabular}{r r r r r r r r r r r r r r}
\hline
                                      & \multicolumn{2}{c}{${\bf 5}_H$} & \multicolumn{3}{c}{${\bf 10}_S$} & \multicolumn{4}{c}{${\bf 35}_S$} \\
$b_{iI}-b_{jI}$ ~\vline    & H                        & $S_1^*$ ~\vline          & $\phi_1$     & $\rtwo{10}$ & $\phi_3$ ~\vline        & $\eta_1$ &            $\eta_2$ & $\eta_3$ & $\eta_4$ \\
\hline
$b_{1I}-b_{2I}$ ~\vline & $-\frac{1}{15}$ & $\frac{1}{15}$ ~\vline                      & $\frac{1}{5}$ & $-\frac{7}{15}$             & $\frac{4}{15} ~\vline $ & $\frac{2}{15}$ & $-\frac{6}{5}$ & $-\frac{14}{15}$ & 2 \\
$b_{2I}-b_{3I}$ ~\vline & $\frac{1}{6}$    & $-\frac{1}{6}$  ~\vline & 0                     & $\frac{1}{6}$                & $-\frac{1}{6} ~\vline $  & $\frac{5}{3}$   & $\frac{3}{2}$  & $-\frac{2}{3}$     & $-\frac{15}{6}$ \\
\hline
\end{tabular}
\end{table}
According to Table \ref{table:beta}, the split multiplets that can possibly reduce the value of $B_{12}$ and, in turn, lead to unification at a high value of $M_{GUT}$ are: $\rtwobar~, \eta_2$, and $\eta_3$. The first of these particles, i.e., $\rtwobar$ is a scalar leptoquark. This leptoquark does not lead to proton decay at tree level by itself as it has no diquark couplings. This fact renders the lightness of this particle possible. $\rtwobar$ is restricted to lie in the range $2.4 \times 10^{9}~\text{GeV} \leq m_{\rtwobar} \leq 2.5 \times 10^{15}~\text{GeV}$ as can be seen in Tables \ref{table:masses1} and \ref{table:masses2}. The smallest figure we chose for $m_{\rtwobar}$ is just above the lower bound of $2.1 \times 10^9$ GeV. As the scalar leptoquark $\rtwobar$ has to run in the loop seen in Figure \ref{fig:neutrinomass}, the neutrino mass scale will come out correctly, due to perturbativity constraints, if there's an upper bound on the mass of $\rtwobar$. Whereas the lower limit is necessary to prevent the contribution of the diagram given in Figure \ref{fig:proton decay} to the decay rate of the proton from exceeding the bound allowed by experiments. These limits were obtained in Section \ref{neutrino} on neutrino masses. \par
The other scalar leptoquark is the $S_1^*$ whose mass is constrained to be greater than $m_{S_1^*} \geq 2.8 \times 10^{11}~\text{GeV}$ \cite{Dorsner:2012uz}. We have updated the bound found in that article using the aforementioned experimental figure of $\tau_p (p \rightarrow K^+ {\overline \nu}) > 5.9 \times 10^{33}$ years. Despite its positive contribution to $B_{12}$ which lowers the GUT scale slightly, $S_1^*$ is crucial for nonzero neutrino mass generation through the leptoquark mixing mechanism. The upper bound on the mass of $\rtwobar$ is also valid for $S_1^*$ for the same reason, and we allow the mass of $S_1^*$ to vary in the interval  $2.8 \times 10^{11}~\text{GeV} \leq m_{S_1^*} \leq 2.5 \times 10^{15}~\text{GeV}$ in our unification analysis as can be seen in Tables \ref{table:masses1} and \ref{table:masses2}. \par
Still, it's the contribution of the split multiplets $\eta_2$ and $\eta_3$ from $\scalar{35}$ that plays the dominant role in unification. Among these split multiplets $\eta_3$ prefers to be light, that is why we placed the mass of $\eta_3$, which is the scale of new physics as well, at 1 (10) TeV. Consequently, there are only two independent masses which are that of $\eta_2$ and $\mgut$. The two equations that need to be satisfied can then be solved to find these masses. The values of these masses obtained by accounting for threshold corrections using the two-loop equations are shown for various scenarios in Tables \ref{table:masses1} and \ref{table:masses2}. \par
It's also possible to find a general formula for the GUT scale and the mass of $\eta_2$ by solving the one-loop equations with the above constraints implemented. This formula, of course, can't be trusted numerically, nevertheless, it captures the general features of the solution that includes two-loop running of coupling constants. This is the content of Eq.(\ref{gutscale}) in which both the GUT scale and the mass of $\eta_2$ is in units of GeV. \par
\begin{eqnarray}
\label{gutscale}
\nonumber \mgut &=& 8.30 \times 10^{15} ~ \left( \frac{m_{\eta_3}}{10~\text{TeV}} \right)^{-22/127} ~ \left( \frac{m_{S_1^*}}{1.2 \times 10^{14}~\text{GeV}} \right)^{-1/127} ~ \left( \frac{m_{\rtwobar}}{2.5 \times 10^{15}~\text{GeV}} \right)^{-5/127} \\
m_{\eta_2} &=& 4.76 \times 10^{7} ~ \left( \frac{m_{\eta_3}}{10~\text{TeV}} \right)^{-36/381} ~ \left( \frac{m_{S_1^*}}{1.2 \times 10^{14}~\text{GeV}} \right)^{+33/381} ~ \left( \frac{m_{\rtwobar}}{2.5 \times 10^{15}~\text{GeV}} \right)^{-89/381}
\end{eqnarray}
\begin{table}
\centering
\caption{Two-loop analysis results for the mass of $ \eta_2$ and the unification scale for selected masses of $S_1^*$ and $\rtwobar$ in the case in which the mass of $\eta_3$ is at 1 TeV.}
\label{table:masses1}
\begin{tabular}{c c c c c }
\hline
$m_{S_1^*}$~[GeV] & $m_{\rtwobar}$~[GeV] & $m_{\eta_2}$~[$10^{10}$~GeV] & $\mgut$~[$10^{16}$~GeV] \\
\multirow{3}{*}{$9.2 \times 10^{13}$} & $2.4 \times 10^9$ & 10.9 & 4.53 \\
                                                                                                       & $1.9 \times 10^{10}$ & 6.77 & 4.14 \\
									       & $1.3 \times 10^{12}$ & 2.54 & 3.45 \\
\hline
\multirow{4}{*}{$2.5 \times 10^{15}$} & $2.4 \times 10^9$ & 14.2 & 4.40 \\
                                                                                                & $1.5 \times 10^{10}$ & 9.31 & 4.07 \\
                                                                                                & $8.4 \times 10^{11}$ & 3.65 & 3.41 \\
                                                                                                & $4.6 \times 10^{13}$ & 1.46 & 2.87 \\
\hline
$2.8 \times 10^{11}$ & \multirow{2}{*}{$4.6 \times 10^{13}$} & 0.710 & 3.11 \\
$3.6 \times 10^{12}$ & & 0.869 & 3.04 \\
\hline
$2.8 \times 10^{11}$ & \multirow{3}{*}{$2.5 \times 10^{15}$} & 0.288 & 2.62 \\
$5.8 \times 10^{12}$ & & 0.365 & 2.55 \\
$1.2 \times 10^{14}$ & & 0.464 & 2.49 \\
\hline
\end{tabular}

\end{table}
\begin{table}
\centering
\caption{Two-loop analysis results for the mass of $ \eta_2$ and the unification scale for selected masses of $S_1^*$ and $\rtwobar$ in the case in which the mass of $\eta_3$ is at 10 TeV.}
\label{table:masses2}
\begin{tabular}{c c c c c }
\hline
$m_{S_1^*}$~[GeV] & $m_{\rtwobar}$~[GeV] & $m_{\eta_2}$~[$10^{10}$~GeV] & $\mgut$~[$10^{16}$~GeV] \\
\multirow{3}{*}{$9.2 \times 10^{13}$} & $2.4 \times 10^9$ & 7.27 & 2.58 \\
                                                                                                       & $1.9 \times 10^{10}$ & 4.50 & 2.36 \\
									       & $1.3 \times 10^{12}$ & 1.69 & 1.96 \\
\hline
\multirow{4}{*}{$2.5 \times 10^{15}$} & $2.4 \times 10^9$ & 9.47 & 2.51 \\
                                                                                                & $1.5 \times 10^{10}$ & 6.20 & 2.32 \\
                                                                                                & $8.4 \times 10^{11}$ & 2.43 & 1.94 \\
                                                                                                & $4.6 \times 10^{13}$ & 0.970 & 1.64 \\
\hline
$2.8 \times 10^{11}$ & \multirow{2}{*}{$4.6 \times 10^{13}$} & 0.473 & 1.77 \\
$3.6 \times 10^{12}$ & & 0.579 & 1.73 \\
\hline
$2.8 \times 10^{11}$ & \multirow{3}{*}{$2.5 \times 10^{15}$} & 0.192 & 1.50 \\
$5.8 \times 10^{12}$ & & 0.243 & 1.46 \\
$1.2 \times 10^{14}$ & & 0.310 & 1.42 \\
\hline
\end{tabular}
\end{table}
According to Eq.(\ref{gutscale}), decreases in the masses of $\eta_3$ and $\rtwobar$ both lead to a higher GUT scale, due to their negative contributions to the $B_{12}$ coefficient, and a heavier $\eta_2$. These trends can be seen to hold in the solutions which include threshold corrections in Tables \ref{table:masses1} and \ref{table:masses2}, although the exponents of the terms in parantheses in Eq.(\ref{gutscale}) can't be expected to be correct. Likewise, the lowest value of the GUT scale predicted to be $8.30 \times 10^{15}~\text{GeV}$ by Eq.(\ref{gutscale}) underestimates that stated in Table \ref{table:masses2} which is $1.42 \times 10^{16} ~\text{GeV}$. On the other hand, one would think that an increase in the mass of the scalar leptoquark $S_1^*$ would increase the GUT scale in view of its positive contribution to the $B_{12}$ coefficient, however, it should be borne in mind that as the mass of this particle changes so does the mass of $\eta_2$ for unification to occur and that increase eclipses the effect of $S_1^*$ on the GUT scale. \par
\begin{figure}[h]
\centering
\includegraphics{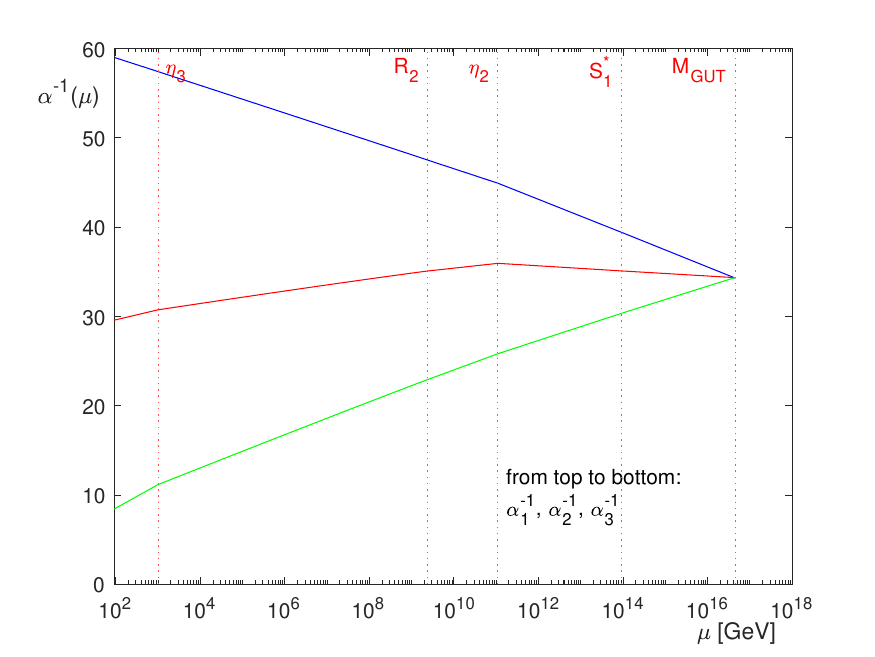}
\caption{Unification scenario with the highest GUT scale of $\mgut=4.53 \times 10^{16}$ GeV in which the split multiplets are at $m_{\eta_3}=1$ TeV, $m_{S_1^*}=9.2 \times 10^{13}$ GeV, $m_{\rtwobar}=2.4 \times 10^9$ GeV and $m_{\eta_2}=1.09 \times 10^{11}$ GeV.}
\label{fig:unification1}
\end{figure}
\begin{figure}[h]
\centering
\includegraphics{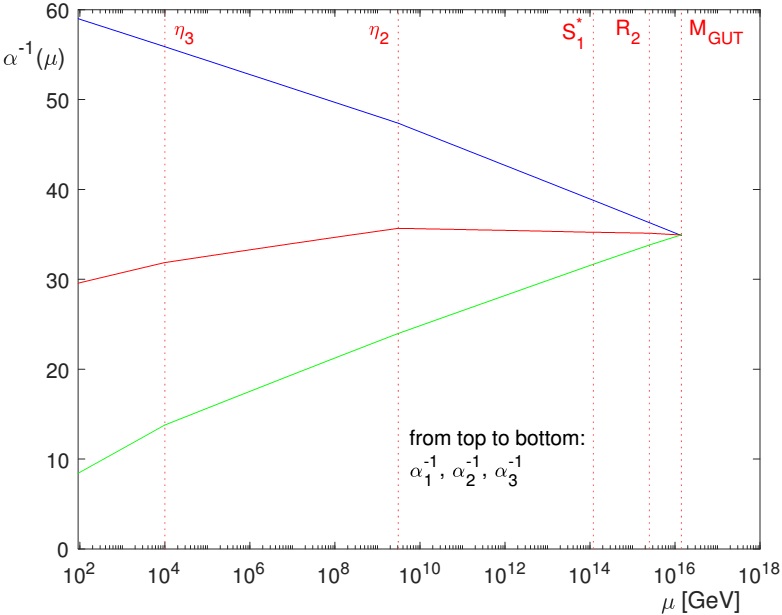}
\caption{Unification scenario with the lowest GUT scale of $\mgut=1.42 \times 10^{16}$ GeV in which the split multiplets are at $m_{\eta_3}=10$ TeV, $m_{S_1^*}=1.2 \times 10^{14}$ GeV, $m_{\rtwobar}=2.5 \times 10^{15}$ GeV and $m_{\eta_2}=3.10 \times 10^9$ GeV.}
\label{fig:unification2}
\end{figure}
We have plotted the variation of the inverse of the three coupling constants with energy scale in units of GeV for two selected scenarios in Figures \ref{fig:unification1} and \ref{fig:unification2}. In Figure \ref{fig:unification1}, the scale of new physics is chosen to be 1 TeV, while the graph in Figure \ref{fig:unification2} corresponds to a scale of 10 TeV for new physics discoveries. The masses of $S_1^*$ and $\rtwobar$ are at $9.2 \times 10^{13}$ GeV and $2.4 \times 10^{9}$ GeV in the first figure, and equal to $1.2 \times 10^{14}$ GeV and $2.5 \times 10^{15}$ GeV in the second one. \par
At this point, we would like to comment on the interactions of the light scalar particle, $\eta_3$, predicted by our model. It has no direct couplings to any of the fermions in the SM. In this regard, it's quite different from the sextet scalars discussed in the literature which couple to SM fermions at tree level and as a result are either weak isosinglets or weak isotriplets see, for example, \cite{Patel:2022wya}. The operator of lowest dimension connecting $\eta_3$ to SM fermions is of dimension 5: ${\overline {\bf 5}}_{Fi} {\bf 5}^C_{Fj} \scalar{35}^{ijk} {\bf 5}^*_{Hk}$. Family indices are suppressed for simplicity.

\section{Charged fermion masses}\label{chargedfermion}
In the absence of the $\higgs{45}$ representation which comprises the second Higgs doublet, one is left with nonrenormalizable operators \cite{Ellis:1979fg} to modify the undesirable equality at the GUT scale of the masses of the d-type quarks and the charged leptons in each of the lighter two generations, i.e., $m_d(\mgut)=m_e(\mgut)$ and $m_s(\mgut)=m_\mu(\mgut)$. The cutoff scale, $\Lambda$, appearing in the nonrenormalizable terms and the ensuing mass matrices of the charged fermions lies in the interval $M_{GUT} \ll \Lambda \leq M_{\text{Pl}}$ where $M_{\text{Pl}}$ is the Planck mass. Yukawa coupling matrix $Y^U$ that determines the mass of the u-type quarks and those which determine the masses of the d-type quarks and charged leptons, namely $Y^D, {\widetilde Y}^D,$ and ${\overline Y}^D$ are 3$\times$3 complex matrices in generation space. The Lagrangian ${\mathcal L}_Y$ relevant for charged fermion masses reads:
\begin{eqnarray}
\label{yukawamass}
\nonumber -{\mathcal L}_Y & \supset & \epsilon_{ijklm} {\bf 10}_{Fa}^{ij} Y^U_{ab} {\bf 10}_{Fb}^{kl} {\bf 5}_H^m \\
&+& \left(Y^D_{ab} {\bf 10}^{ij}_{Fa} + \frac{{\widetilde Y}^D_{ab}}{\Lambda} {\bf 10}^{ik}_{Fa} {\bf 24}_{Hk}^j + \frac{{\overline Y}^D_{ab}}{\Lambda} {\bf 10}^{kj}_{Fa} {\bf 24}_{Hk}^i \right) {\overline {\bf 5}}_{Fbi} {\bf 5}^*_{Hj} + H.c.
\end{eqnarray}
where subscripts $a,b=1,2,3$ of Yukawa coupling matrices denote generation indices. Lagrangian given in Eq.(\ref{yukawamass}) leads to the following mass matrices $M^D, M^E,$ and $M^U$ of the d-type quarks, charged leptons, and u-type quarks, respectively:
\begin{eqnarray}
\nonumber M^D &=& \left[ \frac{1}{2} Y^D - \frac{3}{2} \left(\frac{v_{24}}{\Lambda \sqrt{30}}\right) \left({\widetilde Y}^D - \frac{2}{3} {\overline Y}^D \right) \right]v_5^*, \\
\nonumber \left( M^E \right)^T &=&  \left[ \frac{1}{2} Y^D - \frac{3}{2} \left(\frac{v_{24}}{\Lambda \sqrt{30}}\right) \left({\widetilde Y}^D + {\overline Y}^D \right) \right]v_5^*, \\
M^U &=& \sqrt{2} \left[ Y^U+\left(Y^U\right)^T\right] v_5. 
\end{eqnarray}
The difference in the coefficients of the terms proportional to ${\overline Y}^D$ in the mass matrices for down-type quarks and charged leptons causes these to differ from one another and thus modifies the ratios of the masses of the two lighter generations to agree with the observed values at low energies.  

\section{Conclusion}\label{conclusion}
We have investigated a non-supersymmetric, non-renormalizable SU(5) grand unification model that extends the particle content of GG model in the minimal manner by the following scalars only: $\scalar{10}$ and $\scalar{35}$. We could achieve unification of gauge couplings at the two-loop level at values of the unification scale satisfying $\mgut \geq 1.42 \times 10^{16}$ GeV. Masses of scalar leptoquarks $S_1^*$ and $\rtwobar$ were chosen accordingly in order to generate the elements of the Majorana mass matrix of neutrinos. $\eta_3$ particle which came out to be light as a result of unification constraints was placed at 1 (10) TeV. The remaining two mass scales $\mgut$ and the mass of $\eta_2$ were found numerically by solving the coupled differential equations involving the two-loop beta functions. In order to generate the neutrino Majorana mass matrix with perturbative Yukawa couplings and the leptoquark-Higgs coupling implied a common upper bound on the masses of the leptoquarks as $m_{S^*_1},~m_{\rtwobar} \leq 2.5 \times 10^{15}$ GeV. This calculated figure refines the estimates made in previous work. Unlike the scalar leptoquark $S_1^*$, $\rtwobar$ doesn't have diquark couplings, so its mass was seemingly not constrained from below. Nevertheless, it was found that a too light $\rtwobar$, due to its mixing with the other leptoquark, would lead to a proton lifetime too short to be in compliance with the experimental bounds, were it to fall below $m_{\rtwobar} \geq 2.1 \times 10^9$ GeV. With these constraints implemented we could reproduce the elements of the neutrino mass matrix in both normal and inverted types of hierarchical neutrino mass spectrum by choosing the Yukawa couplings and leptoquark-Higgs coupling in the interval $0.1 \leq Y^\nu, Y^D, \lambda_{5-10} \leq \sqrt{4\pi}$. Lastly, we could generate the mass matrices for charged fermions correctly \`a la Ellis and Gaillard through the nonrenormalizable modifications to the down-type quark and charged lepton mass matrices. We intend to analyze the phenomenology of the predicted sextet scalar particle $\eta_3$ by writing down its couplings to SM particles and calculating the resulting cross-sections in future research.

\appendix
\numberwithin{equation}{section}
\makeatletter
\def\@seccntformat#1{\csname Pref@#1\endcsname \csname the#1\endcsname\quad}
\def\Pref@section{Appendix~}
\makeatother
\include{appendix_a}
\section{Two-loop running of gauge couplings} \label{appendix:twoloop}
At the two-loop level, we justify the identification of the mass of the heavy gauge bosons with the GUT scale, i.e., $M_V=\mgut$ by implementing the following boundary conditions at the GUT scale as was shown in \cite{Hall:1980kf}:
\begin{equation}
\alphagut^{-1}\left(\mgut\right) = \alpha_1^{-1}\left(\mgut\right)+\frac{5}{12\pi} = \alpha_2^{-1}\left(\mgut\right)+\frac{3}{12\pi} = \alpha_3^{-1}\left(\mgut\right)+\frac{2}{12\pi}
\end{equation}
and the (coupled) differential equations obeyed by the gauge couplings are:
\begin{equation}
\label{couplingflow}
\frac{d\alpha_i(\mu)}{d{\text{log}}(\mu)} = \frac{b_i}{2\pi} \alpha_i^2(\mu) + \frac{1}{8\pi^2} \sum_{j=1}^3 b_{ij} \alpha_i^2(\mu) \alpha_j(\mu) + \frac{1}{32\pi^3} \alpha_i^2(\mu) \sum_{l=U,D,E} C_{il} {\text{Tr}}\left[\left(Y^l\right)^\dagger Y^l \right]. 
\end{equation}
Repeated $j,l$ indices on the RHS of Eq.(\ref{couplingflow}) are summed over as usual, but $i$ indices are not! At two-loop order Yukawa couplings contribute, but quartic couplings don't. In fact, one needs to account for the one-loop running of the Yukawa couplings as we show in Appendix \ref{appendix:oneloop}. In order to be able to solve Eq.(\ref{couplingflow}), knowledge of both $b_i$ and $b_{ij}$ coefficients are required. For the SM case $b_i$ and $b_{ij}$ coefficients are listed in \cite{Jones:1981we}. The $b_i$ and $b_{ij}$ coefficients of beyond the SM scalars that figure into our calculation are given below:
\begin{eqnarray}
b_i^{\eta_3} &=& \begin{pmatrix} \frac{1}{15} \\ 1 \\ \frac{5}{3} \end{pmatrix}~,~~~~~~b_i^{\eta_2} = \begin{pmatrix} \frac{4}{5} \\ 2 \\ \frac{1}{2} \end{pmatrix}~,~~~~~~b_i^{S_1^*} = \begin{pmatrix} \frac{1}{15} \\ 0 \\ \frac{1}{6} \end{pmatrix}~,~~~~~~b_i^{\rtwobar} = \begin{pmatrix} \frac{1}{30} \\ \frac{1}{2} \\ \frac{1}{3} \end{pmatrix}~; \\
\nonumber b_{ij}^{\eta_3} &=& \begin{pmatrix} \frac{1}{75} & \frac{3}{5} & \frac{8}{3} \\ \frac{1}{5} & 13 & 40 \\ \frac{1}{3} & 15 & \frac{230}{3} \end{pmatrix}~,~~~~~~b_{ij}^{\eta_2} = \begin{pmatrix} \frac{64}{25} & \frac{96}{5} & \frac{64}{5} \\ \frac{32}{5} & 56 & 32 \\ \frac{8}{5} & 12 & 11 \end{pmatrix}~,~~~~~~b_{ij}^{S_1^*} = \begin{pmatrix} \frac{4}{75} & 0 & \frac{16}{15} \\ 0 & 0 & 0 \\ \frac{2}{15} & 0 & \frac{11}{3} \end{pmatrix}~,~~~~~~ \\
b_{ij}^{\rtwobar} &=& \begin{pmatrix} \frac{1}{150} & \frac{3}{10} & \frac{8}{15} \\ \frac{1}{10} & \frac{13}{2} & 8 \\ \frac{1}{15} & 3 & \frac{22}{3} \end{pmatrix}~.
\end{eqnarray}
These are incorporated at the masses of the respective split multiplets. The $C_{il}$ coefficients are as given in \cite{Arason:1991ic}:
\begin{eqnarray}
C_{il}=
\begin{pmatrix}
\frac{17}{10} & \frac{1}{2} & \frac{3}{2} \\ \frac{3}{2} & \frac{3}{2} & \frac{1}{2} \\ 2 & 2 & 0
\end{pmatrix}~.
\end{eqnarray}

\section{One-loop running of Yukawa couplings} \label{appendix:oneloop}
$\beta^{\text{SM}}_U, \beta^{\text{SM}}_D,$ and $\beta^{\text{SM}}_E$ in Eq.(\ref{yukawarunning}) denote the lowest order beta functions in the SM for up-type quark, down-type quark, and charged lepton Yukawa coupling matrices, respectively \cite{Babu:1987im}.
\begin{eqnarray} \label{yukawarunning}
\nonumber \beta^{\text{SM}}_U &=& T - G^U + \frac{3}{2} \left[\left(Y^U\right)^\dagger Y^U - \left(Y^D\right)^\dagger Y^D \right], \\
\nonumber \beta^{\text{SM}}_D &=& T - G^D + \frac{3}{2} \left[\left(Y^D\right)^\dagger Y^D - \left(Y^U\right)^\dagger Y^U \right], \\
\beta^{\text{SM}}_E &=& T - G^E + \frac{3}{2} \left(Y^E\right)^\dagger Y^E
\end{eqnarray}
where
\begin{equation}
T = {\text{Tr}} \left[ \left(Y^E\right)^\dagger Y^E +  3\left(Y^U\right)^\dagger Y^U +  3\left(Y^D\right)^\dagger Y^D \right]
\end{equation}
and
\begin{eqnarray}
\begin{pmatrix}
G^U \\ G^D \\ G^E
\end{pmatrix}
=
\begin{pmatrix}
\frac{17}{20} & \frac{9}{4} & 8 \\ \frac{1}{4} & \frac{9}{4} & 8 \\ \frac{9}{4} & \frac{9}{4} & 0
\end{pmatrix}
\begin{pmatrix}
g_1^2 \\ g_2^2 \\ g_3^2
\end{pmatrix}~.
\end{eqnarray}
We define the following matrices:
\begin{equation}
M_E \equiv \left(Y^E\right)^\dagger Y^E~,~~~~~~M_D \equiv \left(Y^D\right)^\dagger Y^D~,~~~~~~M_U \equiv \left(Y^U\right)^\dagger Y^U
\end{equation}
whose diagonalization gives
\begin{eqnarray} \label{diagonalization}
\nonumber E^\dagger M_E E &=& {\text{diag}} \left( Y_e^2,Y_\mu^2,Y_\tau^2\right)~~, \\
\nonumber D^\dagger_C M_D D_C &=& {\text{diag}} \left( Y_d^2,Y_s^2,Y_b^2\right)~~, \\
U^\dagger_C M_U U_C &=& {\text{diag}} \left( Y_u^2,Y_c^2,Y_t^2\right)~~.
\end{eqnarray}
The (unitary) matrices $E, D_C,$ and $U_C$ that diagonalize the $M_E,M_D,$ and $M_U$ matrices in Eq.(\ref{diagonalization}), respectively, satisfy the following equations: $U_C^\dagger U_C=D_C^\dagger D_C=E^\dagger E=\mathbb{1}$. CKM matrix is defined in the usual way as follows: $V^{\text{CKM}} = U_C^\dagger D_C$. One can derive the renormalization group equations given below for the diagonal elements of matrices $M_U,M_D,$ and $M_E$, respectively:
\begin{eqnarray}
\nonumber \frac{d\alpha_i^U}{d{\text{log}}\mu} &=& \frac{\alpha_i^U}{2\pi} \left( {\overline T} - {\overline G}^U + \frac{3}{2} \alpha_i^U - \frac{3}{2} \sum_j \left|V^{\text{CKM}}_{ij}\right|^2 \alpha_j^D \right)~, \\
\nonumber \frac{d\alpha_i^D}{d{\text{log}}\mu} &=& \frac{\alpha_i^D}{2\pi} \left( {\overline T} - {\overline G}^D + \frac{3}{2} \alpha_i^D - \frac{3}{2} \sum_j \left|V^{\text{CKM}}_{ji}\right|^2 \alpha_j^U \right)~, \\
\frac{d\alpha_i^E}{d{\text{log}}\mu} &=& \frac{\alpha_i^E}{2\pi} \left( {\overline T} - {\overline G}^E + \frac{3}{2} \alpha_i^E \right)~
\end{eqnarray}
where ${\overline T} \equiv T/\left(4\pi\right), {\overline G}^l \equiv G^l/\left(4\pi\right)$ and $\alpha_i^l \equiv \left(Y_i^l\right)^2/\left(4\pi\right)$ with $i=1,2,3$ being the family index and $l=U,D,E$.

\section{EW mixing of states of Q$_{\mathrm{em}}$=2/3} \label{appendix:leptoquark}
The terms among those in ${\mathcal L}_{\text{mixed}}$ given in Eq.(\ref{lmixed}) that lead to mixing between the particles $\phi_3^*, \rtwobar^{+2/3}$, and $\eta_2^{+2/3}$ of electric charge $\charge =2/3$ is given below.
\begin{eqnarray}
\nonumber \mathcal{L}_{\text{mixed}} & \supset & \epsilon_{ijklm} \left( g_{10-10} {\bf 10}_S^{ij} {\bf 10}_S^{kl} {\bf 5}_H^{m} + \lambda_{10-10} {\bf 10}_S^{ij} {\bf 10}_S^{kl} {\bf 5}_H^{n} {\bf 24}_{Hn}^{m} + {\widetilde \lambda}_{10-10} {\bf 10}_S^{in} {\bf 10}_S^{kl} {\bf 5}_H^{m} {\bf 24}_{Hn}^{j} \right) \\
&+& \lambda_{10-35} {\bf 35}_S^{ijk} {\bf 10}^*_{Sil} {\bf 5}^*_{Hk} {\bf 24}_{Hj}^{l} + H.c.
\end{eqnarray}
The relation given in Eq.(\ref{rotationtwo}) between the mass eigenstates after EW SSB, namely $\rtwohat^{+2/3}, {\widehat \phi}_3^*$, and ${\widehat \eta}_2^{+2/3}$, and the interaction eigenstates which are $\rtwobar^{+2/3}, \phi_3^*$ and $\eta_2^{+2/3}$ defines the rotation matrix, $R^{+2/3}$.
\begin{eqnarray} \label{rotationtwo}
\begin{pmatrix} |{\widehat \phi}_3^*> \\ |\rtwohat^{+2/3}> \\ |{\widehat \eta_2^{+2/3}}> \end{pmatrix} & \equiv & R^{+2/3} \begin{pmatrix} |\phi_3^*> \\ |\rtwobar^{+2/3}> \\ |\eta_2^{+2/3}> \end{pmatrix}~.
\end{eqnarray}
In the basis of interaction eigenstates the squared mass matrix reads
\begin{eqnarray} \label{massmatrixtwo}
{\widetilde {\mathcal M}}^2_{-2/3} &=& \begin{pmatrix}
m_{\phi_3^*}^2 & {\widetilde M}_{12}^2 & 0 \\ \cdot & m_{\rtwobar}^2 & {\widetilde M}_{23}^2 \\ \cdot & \cdot & m_{\eta_2}^2
\end{pmatrix}~,
\end{eqnarray}
while its elements are given below. Again, we display only the elements on or above the diagonal, because the squared mass matrix is symmetric.
\begin{eqnarray} \label{massmatrixeltwo}
\nonumber {\widetilde M}_{12}^2 &=& \frac{v_5 v_{24}}{2\sqrt{15}} \left( \frac{4g_{10-10}}{v_{24}\sqrt{30}} + 3\lambda_{10-10} - 12{\widetilde \lambda}_{10-10} \right)~, \\
{\widetilde M}_{23}^2 &=& \frac{{\overline M}_{23}^2}{\sqrt{2}}~.
\end{eqnarray}
The coupling constants that appear in Eq.(\ref{massmatrixeltwo}), i.e., $g_{10-10}, \lambda_{10-10}~{\text{and}}~{\widetilde \lambda}_{10-10}$ can be taken to be real without loss of generality. The rotation matrix, $R^{+2/3}$, is then found analytically up to corrections of ${\mathcal O}\left(\left({\widetilde M}_{ij}^2/\left(m_i^2-m_j^2\right)\right)^2\right)$ ($i \neq j$) as given below:
\begin{eqnarray}
R^{+2/3} = \begin{pmatrix} 1 & \frac{{\widetilde M}_{12}^2}{\left(m_{\phi_3^*}^2-m_{\rtwobar}^2\right)} & 0 \\ -\frac{{\widetilde M}_{12}^2}{\left(m_{\phi_3^*}^2-m_{\rtwobar}^2\right)} & 1 & \frac{{\widetilde M}_{23}^2}{\left(m_{\rtwobar}^2-m^2_{\eta_2}\right)} \\ 0 & -\frac{{\widetilde M}_{23}^2}{\left(m_{\rtwobar}^2-m^2_{\eta_2}\right)} & 1 \end{pmatrix}~.
\end{eqnarray}

\end{document}